\documentclass[pra,twocolumn,groupedaddress,]{revtex4}
\usepackage{newpxtext,newpxmath}
\usepackage{soul}
\usepackage{sfmath}
\usepackage{amsmath}
\usepackage{graphics}
\usepackage{enumerate}
\usepackage{xcolor}
\usepackage[unicode=true,pdfusetitle,
 bookmarks=true,bookmarksnumbered=false,bookmarksopen=false,
 breaklinks=false,pdfborder={0 0 0},backref=false,colorlinks=true,citecolor=blue]{hyperref}
\usepackage{bbm}
\usepackage{amsmath}
\usepackage[utf8]{inputenc}
\begin{document}

\title{Interference Visibility and Wave-Particle Duality in Multi-Path Interference}\thanks{ \href{http://dx.doi.org/10.1103/PhysRevA.100.042105}{\large\bf Phys. Rev. A 100, 042105 (2019).}}
\author{Tabish Qureshi}
\email{tabish@ctp-jamia.res.in}
\affiliation{Centre for Theoretical Physics, Jamia Millia Islamia, New Delhi,
India.}


\begin{abstract}
Wave-particle duality in multi-path interference is fraught with issues
despite substantial progress in recent years.
It was experimentally shown that in certain specific conditions,
getting path information in a multi-path experiment can actually increase
the visibility of interference. As a result, it was argued that in
multi-path interference experiments,
visibility of interference and ‘which-path’ information are not
always complementary observables. In the present work,
a new wave-particle duality relation
is presented, based on a sum of visibilities of interference from individual
{\em pairs of path}. This relation is always respected, even in the kind
of specific situations mentioned above. This sum of
visibilities turns out to be related to a recently introduced measure of
coherence. As one of the consequences, it provides a novel way of
experimentally measuring coherence in a multi-path interference experiment.
As another consequence, this relation suggests a simple way of measuring
path-distinguishability in multi-path interference. In addition, it 
resolves several outstanding issues concerning wave-particle duality in
multi-path interference.
\end{abstract}

\maketitle

\section{Introduction}

Last two decades have seen lot of research activity in the area of
complementarity or wave-particle duality in multi-path interference
\cite{durr,bimonte,mei,luis,bimonte1,englertmb,prillwitz,3slit,cd,nslit,predict,coles1,bagan,biswas,tania,anu,misba}.
After Englert derived a duality relation $\mathcal{D}^2+\mathcal{V}^2\le 1$,
for a two-path interference, which puts a bound on how much path information
can be obtained from a
quanton and the sharpness of interference it can show \cite{englert}, it was
natural to look for a similar duality relation for multi-slit interference.
Breakthrough came with the derivation of the duality relation 
$\mathcal{D}_Q+\mathcal{C}\le 1$ \cite{cd}, between a new
path-distinguishability $\mathcal{D}_Q$ based on unambiguous quantum state
discrimination (UQSD) \cite{uqsd}, and a new quantity quantum coherence
$\mathcal{C}$, based on quantification of coherence by Baumgratz, Cramer
and Plenio \cite{coherence}.

Despite this tremendous progress, several issues still remained. One was,
how coherence $\mathcal{C}$, which is just based on the $l_1$-norm of the
off-diagonal elements of the density matrix of the quanton, can be measured
in an experiment. A way to measure $\mathcal{C}$ from interference has
been suggested \cite{tania}, but that does not work in all scenarios,
especially the kind
described in the following. Mei and Weitz carried out multi-path 
interference experiments where a controlled decoherence was introduced
only in selected paths \cite{mei}. In addition, the phase of one of the
paths was flipped by $\pi$. In such a situation they saw that increasing
decoherence, which could also amount to getting path information, actually
increased the visibility or contrast of the interference. The visibility
$\mathcal{V}$, it may be recalled, is simply Michelson's expression for
fringe contrast $\mathcal{V} = \frac{I_{max}-I_{min}}{I_{max}+I_{min}}$,
where $I_{max}, I_{min}$ refer to the maximum and minimum intensities
of interference, respectively \cite{bornw}. This is in clear contradiction with the
spirit of the Bohr's principle of complementarity \cite{bohr}. Based on
this result, several authors argued that the interference visibility is
not a good measure of interference or wave nature \cite{bimonte1,luis}.
It was even argued that there exist path measurements which do not
degrade interference \cite{luis}. In this kind of a scenario,
coherence $\mathcal{C}$ can be shown to always decrease with increasing
decoherence, and appears to capture the wave nature of a quanton well.
However, in such a scenario $\mathcal{C}$ cannot be measured from
interference by the method suggested in Ref. \cite{tania}. Thus, it remains an
open question whether a measure of the wave nature can be gotten from 
interference in a multi-path experiment \cite{coh-rev}.
The main result of this paper is the following duality relation for
an unbiased n-path interference
\begin{equation}
\tfrac{2}{n(n-1)}\sum_{\text{pairs}} {\mathcal{D}_Q}_{ij} +
\tfrac{2}{n(n-1)}\sum_{\text{pairs}} \mathcal{V}_{ij} \le 1 ,
\end{equation}
where $\mathcal{V}_{ij}$ is the interference visibility if only the i'th and
j'th slits are open, and the rest are blocked, and ${\mathcal{D}_Q}_{ij}$
is the maximum probability of unambiguously distinguishing between the
i'th and j'th paths in such a scenario. It may be useful to recall that 
$n(n-1)/2$ is the total number of slit pairs, making the two terms,
average of two-path distinguishability, and average of two-path visibility,
with the average taken over all path-pairs. It will be shown that this
inequality will hold in all situations, even the one described by Mei
and Weitz's experiments \cite{mei}.
There are several extremely useful
consequences of this result which will also be discussed in the following.

\section{Interference visibility from pairs of paths}

We begin by writing a general pure state of a quanton passing through
a n-slit or a n-path interferometer. If $|\psi_k\rangle$ represent the
state corresponding to the quanton taking the k'th path, the general
state is given by
\begin{equation}
|\Psi\rangle = c_1|\psi_1\rangle + c_2|\psi_2\rangle + 
\dots + c_n|\psi_n\rangle ,
\end{equation}
where $|c_k|^2$ represents the probability of the quanton taking the k'th path.
The states $\{|\psi_i\rangle\}$ can be assumed to form an ortho-normal set,
without loss of generality.
If we are talking about an experiment in which a path-detector is in
place, which attempts to know which path the quanton followed, the
basic requirement of the theory of quantum measurement is that certain
path detector states should get entangled with the states $\{|\psi_i\rangle\}$:
\begin{equation}
|\Psi\rangle = c_1|\psi_1\rangle|d_1\rangle + c_2|\psi_2\rangle|d_2\rangle + 
\dots + c_n|\psi_n\rangle|d_n\rangle ,
\end{equation}
where $\{|d_i\rangle\}$ represent certain normalized states of the path-detector
which may not necessarily be orthogonal to each other. In case they are 
orthogonal to each other, measuring an observable of the path-detector, which
they are eigenstates of, will reveal which path the particle followed, e.g.,
$|\Psi\rangle \xrightarrow{\text{measurement}} |\psi_k\rangle|d_k\rangle$ (say).
The density matrix for the above entangled state, after tracing over the
path-detector states, can be written as 
\begin{equation}
\rho = Tr_d[|\Psi\rangle\langle\Psi|] = \sum_{i=1}^n \sum_{j=1}^n 
c_ic_j^*|\psi_i\rangle\langle\psi_j|\langle d_j|d_i\rangle
\end{equation}
If the quanton were in a mixed state, for some reason, before encountering
the path-detector, a general form of the state would be given by
\begin{equation}
\rho = \sum_{i=1}^n \sum_{j=1}^n \rho_{ij}
|\psi_i\rangle\langle\psi_j|\langle d_j|d_i\rangle.
\end{equation}
In the subsequent discussion we will assume the above to be the general form
of the density operator, and will specify $\rho_{ij}=c_ic_j^*$ for a
pure quanton state.

Let us suppose that we block all the paths except the paths $i,j$. Then
the effective density matrix of the quanton part will look like
\begin{equation}
\rho^{(2)} = \tfrac{1}{\rho_{ii}+\rho_{jj}}\begin{pmatrix}
\rho_{ii} & \rho_{ij} \\
\rho_{ji} &  \rho_{jj}
\end{pmatrix} ,
\label{rho2}
\end{equation}
where the prefactor has been introduced to
renormalize this 2x2 matrix. The actual density matrix
of the quanton will additionally have $\langle d_j|d_i\rangle$ in the
off-diagonal elements. For a two-slit interference, it is well known that
the fringe visibility is given by twice the absolute value of the off-diagonal
matrix elements. Hence we can write the visibility of interference from
slits $i,j$ as
\begin{equation}
\mathcal{V}_{ij} = \frac{2 |\rho_{ij}| |\langle d_j|d_i\rangle|}{\rho_{ii}+\rho_{jj}}.
\label{Vij}
\end{equation}
Since $|d_i\rangle,|d_j\rangle$ are not in general orthogonal, one can do
a UQSD measurement \cite{uqsd} to determine whether the path-detector state is
$|d_i\rangle$ or $|d_j\rangle$. The specific aspect of UQSD measurements
is that if the method succeeds, one can tell for sure if the state is
$|d_i\rangle$ or $|d_j\rangle$. But sometimes the method fails, giving no
result. If two states $|d_i\rangle,|d_j\rangle$ occur with probabilities
$p_1,p_2$, respectively, the {\em optimal} probability of a successful
distinguishing between the two is given by 
$P_{max} = 1 - 2\sqrt{p_1p_2}|\langle d_j|d_i\rangle|$ \cite{uqsd}.
In our two-slit interference, the probability of the state
$|d_i\rangle,|d_j\rangle$ occurring is $\frac{\rho_{ii}}{\rho_{ii}+\rho_{jj}},\frac{\rho_{jj}}{\rho_{ii}+\rho_{jj}}$, respectively.
So the optimal probability of successfully distinguishing between the two
path-detector states is
$P_{max} = 1 - \frac{2\sqrt{\rho_{ii}\rho_{jj}} }{\rho_{ii}+\rho_{jj}}|\langle d_j|d_i\rangle| $.
Consequently this is also the optimal probability of successfully telling
whether the quanton followed path $i$ or $j$. This optimal probability
is what we define our {\em path-distinguishability} as. Thus, the
path-distinguishability for this two-slit interference is given by
\begin{equation}
{\mathcal{D}_Q}_{ij} = 1 - 2\frac{\sqrt{\rho_{ii}\rho_{jj}}}{\rho_{ii}+\rho_{jj}} |\langle d_j|d_i\rangle|.
\label{Dij}
\end{equation}
Using the (\ref{Vij}) and (\ref{Dij}) we can write
\begin{equation}
{\mathcal{V}_{ij} + \mathcal{D}_Q}_{ij} +  2
\frac{\sqrt{\rho_{ii}\rho_{jj}} -|\rho_{ij}|}{\rho_{ii}+\rho_{jj}}|\langle d_j|d_i\rangle| = 1.
\end{equation}
Since the density matrix given by (\ref{rho2}) is positive
semi-definite, one can write $\sqrt{\rho_{ii}\rho_{jj}} -|\rho_{ij}| \ge 0$.
Thus the above equation implies
\begin{equation}
{\mathcal{D}_Q}_{ij} + \mathcal{V}_{ij} \le 1.
\label{DRm}
\end{equation}
This is a wave-particle duality relation for two-path interference \cite{awpd}.
For a pure quanton state, $\rho_{ij}=c_ic_j^*$ leads to
$\sqrt{\rho_{ii}\rho_{jj}} -|\rho_{ij}| = 0$, and the duality relation 
saturates to an equality
\begin{equation}
{ \mathcal{D}_Q}_{ij} + \mathcal{V}_{ij} = 1.
\label{DRp}
\end{equation}
The result is that if all but two slits are closed, the effectively two-slit
interference follows a tight duality relation (\ref{DRm}), which saturates
for the pure case.

This same procedure can be followed for {\em all pairs} of slits, thus
yielding $\mathcal{V}_{ij}$ and ${\mathcal{D}_Q}_{ij}$ for all pairs
$i,j$. Adding (\ref{DRm}) for all pairs of slits, we get
\begin{equation}
\sum_{\text{pairs}} {\mathcal{D}_Q}_{ij} +
\sum_{\text{pairs}} \mathcal{V}_{ij} \le \frac{n(n-1)}{2} ,
\end{equation}
because for $n$ slits, there are $\frac{n(n-1)}{2}$ pairs. Dividing both
sides by $\frac{n(n-1)}{2}$ we get the required duality relation
\begin{equation}
\tfrac{2}{n(n-1)}\sum_{\text{pairs}} {\mathcal{D}_Q}_{ij} +
\tfrac{2}{n(n-1)}\sum_{\text{pairs}} \mathcal{V}_{ij} \le 1 ,
\label{DRsum}
\end{equation}
which, for pure quanton state, will reduce to an equality.
A skeptic may be excused for asking if the above relation, obtained by
selectively opening only one pair of paths at a time, has anything
to do with genuine multi-path interference. After all, we know that even
for a three-slit experiment, the three-slit interference pattern cannot be
obtained simply as a sum of the interference patterns from various 
{\em pairs of slits}.
To address this criticism we delve deeper into (\ref{DRsum}), to understand
its meaning.

\section{Interference visibility and coherence}

We first consider the case where all the paths are equally probably, which
implies that $\rho_{ii}=\frac{1}{n},~i=1,n$. Two-path distinguishability
and visibility, in this situation, are given by
\begin{eqnarray}
{\mathcal{D}_Q}_{ij} &=& 1 - n\sqrt{\rho_{ii}\rho_{jj}} |\langle d_j|d_i\rangle|\nonumber\\
\mathcal{V}_{ij} &=& n |\rho_{ij}| |\langle d_j|d_i\rangle|.
\end{eqnarray}
We substitute these expressions for ${\mathcal{D}_Q}_{ij}$ and
$\mathcal{V}_{ij}$ into (\ref{DRsum}) to get
\begin{eqnarray}
1 - \tfrac{1}{n-1}\sum_{i\ne j} \sqrt{\rho_{ii}\rho_{jj}} |\langle d_j|d_i\rangle|) 
+ \tfrac{1}{n-1}\sum_{i\ne j} |\rho_{ij}| |\langle d_j|d_i\rangle| \le 1 ,
\nonumber\\
\label{DRexp2}
\end{eqnarray}
where we have used the fact that $\sum_{i\ne j} = 2\sum_{\text{pairs}}$.
From an earlier study of wave-particle duality in n-path interference,
we recall \cite{cd}
\begin{eqnarray}
\mathcal{D}_Q &=& 1 - \tfrac{1}{n-1}\sum_{i\ne j} \sqrt{\rho_{ii}\rho_{jj}} |\langle d_j|d_i\rangle|) \nonumber\\
\mathcal{C} &=& \tfrac{1}{n-1}\sum_{i\ne j} |\rho_{ij}| |\langle d_j|d_i\rangle|,
\label{CD}
\end{eqnarray}
where $\mathcal{D}_Q$ is path-distinguishability defined earlier for
n-path interference, and $\mathcal{C}$ is the {\em coherence} defined
again for n-path interference. Using (\ref{CD}), the duality relation
(\ref{DRsum}) assumes the form
\begin{equation}
\mathcal{D}_Q + \mathcal{C} \le 1,
\label{cd}
\end{equation}
which is exactly the duality relation derived in Ref. \cite{cd}.
So, for symmetric multi-path interference, we get a very elegant connection
of the path-distinguishability
and coherence of n-path interference with the path-distinguishability
and {\em visibility} of two-path interference of pairs of slits or paths:
\begin{eqnarray}
\mathcal{D}_Q = \tfrac{2}{n(n-1)}\sum_{\text{pairs}} {\mathcal{D}_Q}_{ij},
~~~~~~~~
\mathcal{C} = \tfrac{2}{n(n-1)}\sum_{\text{pairs}} \mathcal{V}_{ij}.
\label{connect}
\end{eqnarray}
The immense usefulness of this connection will become clear in the following
analysis, which also applies to the case of unequal intensities 
in different paths, which is discussed later.

What (\ref{connect}) implies for coherence is that in an n-path interference
with equal intensities in all beams,
coherence $\mathcal{C}$ can be obtained simply by opening only a pair of
path at a time and measuring visibility by the conventional method, and
then {\em averging} this visibility over all the pairs of paths.
Thus, (\ref{connect}) provides a simple way of directly obtaining coherence
from the interference pattern, although by the special procedure mentioned above.
The other important consequence of (\ref{connect}) is that in the kind
of experiment hooked up by Mei and Weitz \cite{mei}, flipping the phase of
one path by $\pi$ will have no effect on the visibility of interference
from {\em any two} paths, if all other paths are blocked. Thus n-path
coherence can be measured as easily in Mei and Weitz's experiment, as
in any normal multi-path interference. This method then provides good
measure of wave nature of a quanton, which can be obtained from the
interference from various path pairs. Not only is coherence $\mathcal{C}$
a good measure of wave nature, it can be obtained from the interference
in all situations, contrary to the pessimistic view taken by
some authors \cite{luis,bimonte1}.

In earlier studies on wave-particle duality in multi-path interference
\cite{3slit,nslit}, the distinguishability $\mathcal{D}_Q$ is defined
as an upper bound on the probability with which the $n$ paths can be
{\em unambiguously} distinguished from each other. One problem with
this upper bound is that it is not the optimal probability, meaning
there is no guaranty that this limit will be achievable for a give set
of states $\{|d_i\rangle\}$. The second problem with $\mathcal{D}_Q$ is
that UQSD for more than two states works only for a linearly independent
set $\{|d_i\rangle\}$. If the states are linearly {\em dependent},
UQSD cannot be used, and there is no meaning one can assign to the
expression (\ref{CD}) for $\mathcal{D}_Q$.
The relations (\ref{connect}) solve this problem. Even if the states are
linearly dependent, (\ref{connect}) gives a well defined meaning to
$\mathcal{D}_Q$, in terms of the sum distinguishabilities of different
pairs of paths. Two-path distinguishability is based on UQSD involving
only two states, and is the optimal probability of distinguishing the two
states. Therefore, $\mathcal{D}_Q$ as defined by (\ref{connect}) is always
experimentally attainable. The third problem is that UQSD has only been
experimentally demonstrated for two states \cite{uqsdexpt}. No one knows
how to implement it for more than two states. Since the present method 
represents the distinguishability in terms of two-path UQSD, it can
be experimentally implemented.

Next we look at the more general case where all paths 
may not be equally probable. Here, instead
of summing the two-path distinguishabilities and visibilities as done in
(\ref{DRsum}), we multiply the duality relation (\ref{DRm}) for each path-pair 
with the sum of probabilities of the two paths involved, and then sum over
all $i,j~(i\ne j)$:
\begin{equation}
\sum_{i\ne j} (\rho_{ii}+\rho_{jj}){\mathcal{D}_Q}_{ij} +
\sum_{i\ne j} (\rho_{ii}+\rho_{jj})\mathcal{V}_{ij} \le \sum_{i\ne j} \rho_{ii}+\rho_{jj} ,
\label{DRAsum1}
\end{equation}
A new duality relation for the asymmetric multi-path interference can then
be written from the above as
\begin{equation}
\tfrac{1}{(n-1)}\sum_{\text{pairs}} (\rho_{ii}+\rho_{jj}){\mathcal{D}_Q}_{ij} +
\tfrac{1}{(n-1)}\sum_{\text{pairs}} (\rho_{ii}+\rho_{jj})\mathcal{V}_{ij} \le 1.
\label{DRAsum}
\end{equation}
Substituting (\ref{Vij}) and (\ref{Dij}) in the above, 
we again get the known duality relation (\ref{cd}).
So we see that the new duality relation (\ref{DRAsum}), for asymmetric
n-path interference, is the same as (\ref{cd}), with the following connection
\begin{eqnarray}
\mathcal{D}_Q &=& \tfrac{1}{n-1}\sum_{\text{pairs}} (\rho_{ii}+\rho_{jj}){\mathcal{D}_Q}_{ij} \nonumber\\
\mathcal{C} &=& \tfrac{1}{n-1}\sum_{\text{pairs}} (\rho_{ii}+\rho_{jj})\mathcal{V}_{ij}.
\label{aconnect}
\end{eqnarray}
As a consistency check, for all equally probable paths,
$(\rho_{ii}+\rho_{jj})=\frac{2}{n}$, and (\ref{DRAsum}) reduces to
(\ref{DRsum}).

It is clear from the preceding analysis that in a general multi-path
interference, where the paths may not be equally probable, the multi-path
distinguishability and multi-path coherence can be experimentally measured
by carrying out a series of experiments where only one pair of paths is
open, and the visibility of interference is measured. However, here 
one also needs to measure the relative intensity of each path in the
multi-path experiment. One can then use (\ref{aconnect}) to get the coherence
$\mathcal{C}$. Similarly, if one is able to set up an experiment to
measure path-distinguishability of a pair of paths, one can use (\ref{aconnect})
to get the path distinguishability $\mathcal{D}_Q$ for the multi-slit
interference.

\section{Conclusion}

To summarize, we have introduced a new way of studying wave-particle 
duality in multi-path interference, by opening only one pair of paths at
a time, and measuring conventional visibility and using UQSD to measure
the distinguishability ${\mathcal{D}_Q}_{ij}$. The multi-slit path
distinguishability $\mathcal{D}_Q$ and multi-path coherence $\mathcal{C}$
(for symmetric paths)
can then be obtained as average of ${\mathcal{D}_Q}_{ij}$ and
average of ${\mathcal{V}}_{ij}$ over all path pairs, respectively.
For a multi-path interference where the paths may not be equally probable,
the same method works, but the average has to be taken with each term
weighted with the total intensity from the two paths of the pair.
This method resolves various outstanding issues in wave-particle duality
in multi-path interference, which are listed below.\\
(1) A way of measuring {\em coherence} in multi-path interference is provided
which works even for the experiment of Mei and Weitz \cite{mei}, where
interference visibility was shown to increase with increasing path knowledge.\\
(2) The method shows that wave-nature can always be characterized using
interference, and that it is complementary to path information, even in
multi-path interference, contrary to existing belief \cite{luis,bimonte1}.\\
(3) Multi-path coherence has been given a new meaning in terms of 
interference visibilities of path pairs.\\
(4) Path-distinguishability in multi-path interference is given a new
meaning in terms of path distinguishability for a pair
of paths.\\
(5) Path-distinguishability in multi-path interference continues to hold
even in the situation when path-detector states form a linearly {\em dependent}
set.\\
(6) There was no known way to measure path-distinguishability $\mathcal{D}_Q$
in a multi-path interference. A way is provided here.

\end{document}